\documentclass[12pt,preprint]{aastex}

\shortauthors{\"Oberg et al.}

\begin{document}

\title{The effects of snowlines on C/O in planetary atmospheres}

\author{Karin I. \"Oberg\altaffilmark{1}, Ruth Murray-Clay}
\affil{Harvard-Smithsonian Center for Astrophysics, 60 Garden Street, Cambridge, MA 02138, USA}
\email{koberg@cfa.harvard.edu}
\and
\author{ Edwin A. Bergin}
\affil{Department of Astronomy, University of Michigan, Ann Arbor, MI 48109, USA}

\altaffiltext{1}{Hubble Fellow}

\begin{abstract}
The C/O ratio is predicted to regulate the atmospheric chemistry in hot Jupiters. Recent observations suggest that some exo-planets, e.g. Wasp 12-b, have atmospheric C/O ratios substantially different from the solar value of 0.54.  In this paper we present a mechanism that can produce such atmospheric deviations from the stellar C/O ratio.  In protoplanetary disks, different snowlines of oxygen- and carbon-rich ices, especially water and carbon monoxide, will result in systematic variations in the C/O ratio both in the gas and in the condensed phase. In particular, between the H$_2$O and CO snowlines most oxygen is present in icy grains -- the building blocks of planetary cores in the core accretion model -- while most carbon remains in the gas-phase. This region is coincidental with the giant-planet forming zone for a range of observed protoplanetary disks. Based on standard core accretion models of planet formation, gas giants that sweep up most of their atmospheres from disk gas outside of the water snowline will have C/O$\sim$1, while atmospheres significantly contaminated by evaporating planetesimals will have stellar or sub-stellar C/O when formed at the same disk radius. The overall metallicity will also depend on the atmosphere formation mechanism, and exo-planetary atmospheric compositions may therefore provide constraints on where and how a specific planet formed. 
\end{abstract}


\section{Introduction}

Giant planets are commonly detected around main-sequence stars. A handful of gas giants orbiting close to their host stars have been investigated spectroscopically and with narrow-band photometry, resulting in constraints on their atmospheric compositions. The atmospheres of these `hot Jupiters' are predicted to be abundant in H$_2$O, CH$_4$ and CO \citep{Fortney10}, and observations qualitatively confirm these predictions \citep{Richardson06,Tinetti07,Grillmair08,Swain10}. The relative abundances of molecules can potentially be used to characterize atmospheric temperature and pressure structures \citep{Fortney10} as well as to enhance our understanding of atmospheric chemistry under extreme conditions. However, chemical predictions are sensitive to deviations from solar C/O ratios; for a typical hot Jupiter atmosphere at 1100~K and 0.01~bar, an increase in the C/O ratio from the solar 0.54 to unity may result in CH$_4$, C$_2$H$_2$ and CN abundance increases by three orders of magnitude \citep{Lodders09}.  

These extreme effects have recently been used to estimate the C/O ratio in WASP 12-b. Based on comparison between an array model predictions and six narrow band photometry points, \citet{Madhusudhan11} report an atmospheric C/O$\sim$1, despite a solar C/O ratio in the host star and an overall planetary metallicity (C/H) that is at most twice the stellar value, but more likely sub-stellar  \citep[$2\times10^{-5}-1\times10^{-3}$ in WASP 12-b versus the super-solar metallicity of the host star corresponding to C/H$\sim6\times10^{-4}$;][]{Hebb09,Madhusudhan11}. Based on analyzing the observations, \citet{Madhusudhan11} further argue that the spectral features originate in the deep layers of the atmosphere, that day-night energy redistribution is efficient and that a thermal inversion layer deeper than 0.01 bar can be ruled out. If the low C/H and high C/O are confirmed, and the observations do trace the bulk of the atmosphere, this planet's composition cannot be explained by previously suggested mechanisms to enhance the planetary C/O ratio.  These processes---pile-up of carbon-rich grains in the disk, and cold trapping of H$_2$O (and thus a large fraction of the oxygen budget) outside of the snowline \citep{Stevenson88,Kuchner05,Lodders09}---cannot account for an enhanced C/O ratio without also increasing C/H. 

We instead investigate the effects of the different snowlines of a few common volatile species as a mechanism for modifying the planet atmosphere C/O ratio compared to stellar values. In this letter we develop a simple, static model to demonstrate the possible effects of this mechanism (\S2). We discuss possible extensions of the model, especially the inclusion of dynamics, in \S3. In \S4 we explore predictions of planetary atmospheric compositions based on the location of the forming planet with respect to different snowlines and the mechanism through which the planet atmosphere forms.

\section{Model prescription and results}

In the core accretion scenario, planet formation begins with the coagulation of icy grains in the protoplanetary disk.  Grains agglomerate into planetesimals (km sized bodies), which in turn combine to form the planet core. When sufficiently massive, the core accretes a gaseous envelope \citep{Pollack96, Hubickyj05}.  Because the main molecular sources of C and O have different condensation temperatures (Table 1), the relative amounts of C and O differ in grains and in gas at some disk radii. Atmospheric C/O need not, therefore, reflect the average C/O of the disk.  Instead, the C/O ratio of a planetary envelope depends on where in the disk the planet forms, on the fractions of the atmosphere that are accreted from gas and from icy bodies, and on the importance of core dredging, i.e. how isolated the atmosphere is from the core.  In the simplest case, the core and atmosphere are completely isolated from each other, and the atmosphere is built up purely from gas. We therefore begin with only considering gas accretion, and then show how the expected atmosphere composition is modified by adding planetesimal accretion.

Once a core is massive enough to begin runaway accretion of a gas envelope, this accretion most likely happens faster than the planet can migrate due to interactions with the disk.  \citet{DAngelo08} estimate that a planet migrates inward by $<$20\% of its semi-major axis during runaway growth. We therefore assume that the planetary envelope is accreted between the same set of snowlines where accretion started. As a first step, we further assume that grains contributing to the atmosphere come from the same location as the gas (which need not be the case) and that gas and grain compositions are constant between each set of snowlines. Finally we assume that the snowlines are static, which is justified by the long timescales at which disk midplane temperatures change in disks older than 10$^6$ years (when gas giants are proposed to form) compared to the 10$^5$ year timescales of runaway gas-accretion \citep{Lissauer09,Dodson-Robinson09}. Specifically, the temperature structure is set by viscous dissipation in the inner disk and irradiation by the central star in the outer disk \citep{dAlessio98}, and both accretion and stellar luminosity decay on 10$^6$ year timescales at the time of planetary envelope accretion \citep[e.g.][]{Hartmann98,Siess00}. We return to these considerations in \S3.

We estimate the total abundances (grain + gas) of the major O- and C-containing species in typical disks from a combination of ice observations of a protoplanetary disk \citep{Pontoppidan06}, and derived grain compositions in the dense interstellar medium (ISM)(Table 1).  The main O carriers are H$_2$O, CO$_2$ and CO ices, CO gas and silicates and an additional refractory oxygen component \citep{Whittet10}. The main C carriers are CO, CO$_2$, and a range of organics and carbon grains \citep{Draine03}. The evaporation temperature of the latter carbon sources are unknown, and a high evaporation temperature is adopted to prevent this unknown carbon component from influencing the model outcome; if any of this carbon is present in more volatile forms, it will enhance the gas-phase C/O ratio further. The sublimation temperature for silicate grains are set to 1500~K. For all other molecules, we calculate the density-dependent sublimation temperatures  following the prescription of \citet{Hollenbach09} using binding energies of H$_2$O, CO$_2$ and CO of 5800~K, 2000~K and 850~K \citep{Collings04,Fraser01,Aikawa96}. A complication is the observed ease with which H$_2$O can trap other molecules in its ice matrix. It is however difficult to trap more than 5--10\% of the total CO abundance in H$_2$O ice \citep{Fayolle11} and we therefore ignore this process. 

The radii of different snowlines are set by the disk temperature profile. Consistent with the temperature profile derived from the compositions of solar system bodies \citep{Lewis74} and with observations of protoplanetary disks \citep{Andrews05,Andrews07} we adopt a power law profile,

\begin{equation}
T=T_{\rm 0}\times\left(\frac{r}{1\:{\rm AU}}\right)^{-q},
\end{equation}

\noindent where $T_{\rm 0}$ is the temperature at 1~AU and $q$ is the power law index. In a large sample of protoplanetary disks, the average $T_{\rm 0}$ is 200~K, and $q=0.62$ \citep{Andrews07}. Figure 1 displays the C/O in the gas and in grains in the disk midplane as a function of distance from the young star for this average disk profile.  Between the H$_2$O and CO snowlines, the gas-phase C/O ratio increases as O-rich ices condense, with the maximum C/O$\sim$1 reached between the CO$_2$ and CO sublimation lines at 10--40~AU. In the case of completely isolated core and atmosphere accretion, the atmospheric C/O ratios will reflect the gas phase abundances, resulting in C enrichments beyond the H$_2$O snowline. 

The size and position of the disk region where the C/O ratio in the gas reaches unity depend on the disk temperature profile. A more luminous star will heat the disk further, pushing the various snowlines outwards, while the steepness of the disk temperature profile determines the spacing of the different snowlines. Figure 2 compares protoplanetary disk thermal profiles from \cite{Andrews05}, which sample stars with a range in spectral types, with the `typical' disk profile from Fig. 1. In all cases, the gas-phase C/O ratio is enhanced in regions associated with gas-giant formation, i.e. a few to a few tens of AU. Formation of C-rich atmospheres from oxygen depleted gas accretion can therefore operate in most planet forming disks

The high metallicity of giant planets in our own solar system as well as planet formation models suggest that the atmosphere can be significantly polluted by evaporating planetesimals during the late stages of planet formation. The final composition of a planetary envelope then depends not only on the radius at which the atmosphere is accreted, but also on the relative importance of gas and planetesimal accretion during atmosphere build-up. We parameterize these contributions to the relative abundance of an element X as

\begin{equation}
a_{\rm X}=\frac{m_{\rm X/H}^{\rm atm}}{m_{\rm X/H}^{\rm stellar}}=\frac{f_{\rm x, solid}}{f_{\rm s/g}}\frac{M_{\rm solid}}{M_{\rm gas}}+(1-f_{\rm x,solid}), 
\end{equation}

\noindent so that $a_{\rm X}$ is the deviation of X=C or O from stellar. Here, $m_{\rm X/H}^{\rm atm}$ and $m_{\rm X/H}^{\rm stellar}$ are the mass ratios of the element X with respect to H in the planetary envelope and star, respectively,  $f_{\rm x, solid}$ is the fraction of X bound up in solids, $f_{\rm s/g}$ is the mass ratio of solids versus gas in the disk, $M_{\rm solid}$ is the atmospheric mass accreted from solids, and  $M_{\rm gas}$ is the atmospheric mass accreted from gas. In the presented model $f_{\rm s/g}$ is set to 0.01, the observed grain to gas ratio in the ISM. The equation assumes that the amount of H bound up in solids is negligible and that the solids and gas are accreted from the same disk region, i.e. it ignores both planet and planetesimal movement.

The effect of solids evaporating in the atmosphere, whether from accreting icy bodies or from core dredging, on the C/O ratio is illustrated in Fig. 3 for an atmosphere accreted between the CO$_2$ and CO snow lines ($f_{\rm C, solid}=0.35$ and $f_{\rm O, solid}=0.67$ from Table 1). Super-stellar C/O ratios are achieved when $M_{\rm s}/M_{\rm g}<f_{\rm s/g}$, independent of the absolute value of $f_{\rm s/g}$. In Fig. 3, super-stellar C/O ratios are produced when less than 1\% of the total atmosphere mass comes from evaporation of solid material since $f_{\rm s/g}=0.01$. Our parameterization also results in a prediction of the C/H ratio as a function of gas/solid atmosphere accretion. The C/H ratio is sub-stellar when the atmosphere is dominated by gas accretion ($>$99\%) since the amount of H bound up in grains and grain mantles is negligible compared to the gas H$_2$ content, while large amounts of C and O are in the grain mantles at these disk radii. 

\section{Model extensions: disk and planet growth dynamics}

So far we have assumed a static disk, in which the gas and solid compositions are set purely by the evaporation lines of different grain mantle molecules, and where the entire planetary envelope forms at the same location. These assumptions ignore several dynamic processes active during planet formation. First, growing planets likely move within their natal disks during the formation process. While run-away gas accretion is simulated to be fast enough to effectively take place at a single radius, planetesimals may accrete over longer times. The solids that build up the atmosphere through either planetesimal accretion or core dredging may therefore originate at different locations compared to where the gas envelope is accreted.

Second, small solids drift inwards while planets are forming. After crossing a snowline, planetesimals evaporate over a finite length scale, which depends on a combination of the surface evaporation rate and the drift rate, enhancing the gas phase abundance in the evaporating species close to the snowline \citep{Brauer08}. The C/H ratio incorporated into the atmosphere of a planet forming just interior of the snowline may therefore be higher compared to in a stationary disk, even super-stellar. Figure 3 includes predictions for the C/H and C/O ratios for a planet that is forming in a region where the CO/H$_2$ ratio is enhanced by a factor of two due to this kind of evaporation front.

Finally the disk temperature profile is not static, but rather depends on both the evolving luminosity of the accreting star and the disk opacity \citep{Lecar06,Makalkin09}. Therefore, even if a planet remains stationary it may accrete its envelope from gas and solids with evolving compositions. As discussed in \S2 this should be a minor effect during gas accretion, but it may affect the planetesimal accretion process, which occurs over longer timescales. Finally, ice and gas also evolves chemically \citep[e.g.][]{Dodson-Robinson09,Visser11} during the disk life time. This is likely to only have a minor effect on the composition of forming planetary envelopes, however, since the abundances of the most important molecules H$_2$O and CO are predicted to stay the same on 10$^6$ year timescales \citep{Aikawa99,Visser11}. 

 Ideally all these dynamic effects should be included when simulating the build-up of a planetary envelope quantitatively. Without these processes included in the model, we can, however, still predict some atmospheric composition trends based on how and where a planet envelope forms with respect to different snowlines, because of the short timescale of runaway gas accretion compared to most other disk processes, including migration, temperature variations and chemical evolution.

\section{Discussion}

\noindent The model prescription above shows that protoplanetary disk models generally contain zones in the planet forming region where the gas is enhanced in carbon with respect to oxygen. The resulting planetary envelope composition will depend on this gas C/O ratio as well as the amount of icy solids incorporated into the atmosphere through planetesimal accretion or core dredging, and the disk location where these solids form. As observations and our understanding of the relationship between atmosphere spectra and the overall atmosphere composition improve, the combination of C/O and C/H may therefore provide an important probe of a planet's formation history. 

In summary, stellar C/O is expected for planets forming through gravitational instabilities where all material is mixed (reduced C/O and enhanced C/H are possible from late planetesimal accretion), and for planets forming interior to both the water snow-line and the carbon-grain evaporation line, where all volatiles are in the gas phase. Sub-stellar or stellar C/O and super-stellar C/H indicates that a large amount of icy solids 'polluted' the atmosphere following gas accretion. Superstellar C/O and C/H may be due either to gas accretion close to the CO or CO$_2$ snowlines, or indicate that the atmosphere was substantially polluted by carbon-grains that have been postulated to be abundant between the carbon-grain sublimation line and the H$_2$O snow-line. Finally, superstellar C/O and substellar C/H are only consistent with atmosphere formation from mainly gas accretion outside of the water snowline. We note that this last case constrains the location of atmospheric accretion, while for super-stellar C/H, we constrain the formation location of the dominant planetesimal pollutants with respect to the locations of different snowlines in a disk. As both ice abundances and exo-planetary abundances of N-bearing species become better determined, C/N ratios will provide additional constraints on the planetary envelope formation history due to the complementary evaporation histories of C and N-bearing molecules. 

Testing these scenarios relies on well-determined atmospheric compositions, which are difficult to achieve even in our own solar system; the measured C/O value for Jupiter is high, but this is believed to reflect a layered atmosphere rather than a constraint on the bulk composition of the planet's envelope \citep{Atreya02}, since in a cool gas giant, such as Jupiter, much of the O-rich volatile material originally present in the envelope may have condensed \citep{Visscher10}. The super-stellar metallicities of Jupiter and Saturn are better established, and indicate that large amounts of icy solids have been incorporated into their envelopes \citep[e.g.][]{Mousis09}. An overall high C/O ratio is therefore not expected, since accretion of solids outside of the H$_2$O snowline decreases the C/O ratio (Fig. 3). Current observations cannot constrain the bulk C/O ratios of Kuiper belt objects (KBOs). However, our model predicts that KBOs formed beyond $\sim$40 AU will have solar C/O ratios, while those formed interior to $\sim$40 AU and transported outward will have sub-solar C/O ratios.  We note that moons may not have formed at the same temperature as their host planets, complicating their use as chemical tracers.

Because of the poor mixing of volatiles in gas giants in our own Solar System, the compositions of hot Jupiters, where circulation is more efficient \citep{Madhusudhan11}, may be comparatively easier to constrain, despite the technical difficulties involved. Of the handful of spectroscopically observed hot Jupiters, WASP 12b and possibly HD 189733B are candidates to have atmospheres originating from mainly gas accretion beyond both the H$_2$O and CO$_2$ snowlines \citep{Madhusudhan11,Swain09,Swain10}, pending confirmation of the reported C/O ratios and C/H ratios and theoretical confirmation of that the observed compositions are representative of the bulk envelope. In both cases the planets must have migrated from their gas accretion zones beyond the snow-line into their current sub-AU orbits. 

In contrast, the directly imaged young planets observed at 24--68 AU around the A star HR 8799 \citep{Marois08} may have formed close to their present location. Some of the planets are currently located  in the exact region where the gas phase C/O ratio is expected to the largest around an A star (Fig. 2). A discovered high C/O ratio would be evidence for that the planetary envelopes formed through mainly gas accretion close to their current location. In contrast, stellar C/O ratios in the planetary envelopes could be the product of either gravitational collapse or gas and planetesimal
accretion further from the star followed by inward migration. 
Sub-stellar C/O in the envelope would instead be evidence for either gravitational collapse or core accretion, followed by significant planetesimal enrichment interior to the CO snowline (Fig. 3). Spectra of HR 8799b, situated at 68~AU, show a lack of CH$_4$ absorption, possibly indicating a low C/O ratio \citep{Bowler10}.  Information on CO and H$_2$O are however needed to constrain the atmospheric composition further. With effective temperatures of $\sim$800--1100~K \citep{Marois08}, the young planets orbiting HR 8799 are hotter than Jupiter but cooler than hot Jupiters.  Whether spectra for these objects constrain their bulk envelope compositions is uncertain and merits further investigation.

In general, accretion of oxygen depleted gas is a simple way of forming planet atmospheres with high C/O ratios that directly arises from the current planet formation paradigm, consisting of core accretion in a disk with a thermal gradient. To predict atmospheric spectra based on the atmosphere formation mechanism requires, however, both detailed simulations that take into account gas and planetesimal accretion as the planet core moves through the disk, and further exploration of the circumstances under which bulk envelope compositions can be constrained from observable planetary atmospheres.

\acknowledgments

We are grateful for comments from an anonymous referee. Support for KIO is provided by NASA through Hubble Fellowship grant awarded by the Space Telescope Science Institute,  operated by the Association of Universities for Research in Astronomy, Inc., for NASA, under contract NAS 5-26555. EAB acknowledges support from NSF via grant \#1008800

\bibliographystyle{aa}

\begin{thebibliography}{40}

\bibitem[{{Aikawa} {et~al.}(1996){Aikawa}, {Miyama}, {Nakano}, \&
  {Umebayashi}}]{Aikawa96}
{Aikawa}, Y., {Miyama}, S.~M., {Nakano}, T., \& {Umebayashi}, T. 1996, ApJ,
  467, 684

\bibitem[{{Aikawa} {et~al.}(1999){Aikawa}, {Umebayashi}, {Nakano}, \&
  {Miyama}}]{Aikawa99}
{Aikawa}, Y., {Umebayashi}, T., {Nakano}, T., \& {Miyama}, S.~M. 1999, \apj,
  519, 705

\bibitem[{{Andrews} \& {Williams}(2005)}]{Andrews05}
{Andrews}, S.~M. \& {Williams}, J.~P. 2005, ApJ, 631, 1134

\bibitem[{{Andrews} \& {Williams}(2007)}]{Andrews07}
{Andrews}, S.~M. \& {Williams}, J.~P. 2007, ApJ, 659, 705

\bibitem[{{Atreya} {et~al.}(2002){Atreya}, {Mahaffy}, {Niemann}, \&
  {Owen}}]{Atreya02}
{Atreya}, S.~K., {Mahaffy}, P.~R., {Niemann}, H.~B., \& {Owen}, T.~C. 2002,
  Highlights of Astronomy, 12, 597

\bibitem[{{Bowler} {et~al.}(2010){Bowler}, {Liu}, {Dupuy}, \&
  {Cushing}}]{Bowler10}
{Bowler}, B.~P., {Liu}, M.~C., {Dupuy}, T.~J., \& {Cushing}, M.~C. 2010, ApJ,
  723, 850

\bibitem[{{Brauer} {et~al.}(2008){Brauer}, {Henning}, \&
  {Dullemond}}]{Brauer08}
{Brauer}, F., {Henning}, T., \& {Dullemond}, C.~P. 2008, \aap, 487, L1

\bibitem[{{Collings} {et~al.}(2004){Collings}, {Anderson}, {Chen}, {Dever},
  {Viti}, {Williams}, \& {McCoustra}}]{Collings04}
{Collings}, M.~P., {Anderson}, M.~A., {Chen}, R., {et~al.} 2004, MNRAS, 354,
  1133

\bibitem[{{D'Alessio} {et~al.}(1998){D'Alessio}, {Canto}, {Calvet}, \&
  {Lizano}}]{dAlessio98}
{D'Alessio}, P., {Canto}, J., {Calvet}, N., \& {Lizano}, S. 1998, \apj, 500,
  411

\bibitem[{{D'Angelo} \& {Lubow}(2008)}]{DAngelo08}
{D'Angelo}, G. \& {Lubow}, S.~H. 2008, \apj, 685, 560

\bibitem[{{Dodson-Robinson} {et~al.}(2009){Dodson-Robinson}, {Willacy},
  {Bodenheimer}, {Turner}, \& {Beichman}}]{Dodson-Robinson09}
{Dodson-Robinson}, S.~E., {Willacy}, K., {Bodenheimer}, P., {Turner}, N.~J., \&
  {Beichman}, C.~A. 2009, \icarus, 200, 672

\bibitem[{{Draine}(2003)}]{Draine03}
{Draine}, B.~T. 2003, ARA\&A, 41, 241

\bibitem[{{Fayolle} {et~al.}(2011){Fayolle}, {{\"O}berg}, {Cuppen}, {Visser},
  \& {Linnartz}}]{Fayolle11}
{Fayolle}, E.~C., {{\"O}berg}, K.~I., {Cuppen}, H.~M., {Visser}, R., \&
  {Linnartz}, H. 2011, A\&A, 529, A74+

\bibitem[{{Fortney} {et~al.}(2010){Fortney}, {Shabram}, {Showman}, {Lian},
  {Freedman}, {Marley}, \& {Lewis}}]{Fortney10}
{Fortney}, J.~J., {Shabram}, M., {Showman}, A.~P., {et~al.} 2010, ApJ, 709,
  1396

\bibitem[{{Fraser} {et~al.}(2001){Fraser}, {Collings}, {McCoustra}, \&
  {Williams}}]{Fraser01}
{Fraser}, H.~J., {Collings}, M.~P., {McCoustra}, M.~R.~S., \& {Williams}, D.~A.
  2001, MNRAS, 327, 1165

\bibitem[{{Grillmair} {et~al.}(2008){Grillmair}, {Burrows}, {Charbonneau},
  {Armus}, {Stauffer}, {Meadows}, {van Cleve}, {von Braun}, \&
  {Levine}}]{Grillmair08}
{Grillmair}, C.~J., {Burrows}, A., {Charbonneau}, D., {et~al.} 2008, Nature,
  456, 767

\bibitem[{{Hartmann} {et~al.}(1998){Hartmann}, {Calvet}, {Gullbring}, \&
  {D'Alessio}}]{Hartmann98}
{Hartmann}, L., {Calvet}, N., {Gullbring}, E., \& {D'Alessio}, P. 1998, \apj,
  495, 385

\bibitem[{{Hebb} {et~al.}(2009){Hebb}, {Collier-Cameron}, {Loeillet},
  {Pollacco}, {H{\'e}brard}, {Street}, {Bouchy}, {Stempels}, {Moutou},
  {Simpson}, {Udry}, {Joshi}, {West}, {Skillen}, {Wilson}, {McDonald},
  {Gibson}, {Aigrain}, {Anderson}, {Benn}, {Christian}, {Enoch}, {Haswell},
  {Hellier}, {Horne}, {Irwin}, {Lister}, {Maxted}, {Mayor}, {Norton}, {Parley},
  {Pont}, {Queloz}, {Smalley}, \& {Wheatley}}]{Hebb09}
{Hebb}, L., {Collier-Cameron}, A., {Loeillet}, B., {et~al.} 2009, ApJ, 693,
  1920

\bibitem[{{Hollenbach} {et~al.}(2009){Hollenbach}, {Kaufman}, {Bergin}, \&
  {Melnick}}]{Hollenbach09}
{Hollenbach}, D., {Kaufman}, M.~J., {Bergin}, E.~A., \& {Melnick}, G.~J. 2009,
  ApJ, 690, 1497

\bibitem[{{Hubickyj} {et~al.}(2005){Hubickyj}, {Bodenheimer}, \&
  {Lissauer}}]{Hubickyj05}
{Hubickyj}, O., {Bodenheimer}, P., \& {Lissauer}, J.~J. 2005, Icarus, 179, 415

\bibitem[{{Kuchner} \& {Seager}(2005)}]{Kuchner05}
{Kuchner}, M.~J. \& {Seager}, S. 2005, ArXiv Astrophysics e-prints

\bibitem[{{Lecar} {et~al.}(2006){Lecar}, {Podolak}, {Sasselov}, \&
  {Chiang}}]{Lecar06}
{Lecar}, M., {Podolak}, M., {Sasselov}, D., \& {Chiang}, E. 2006, ApJ, 640,
  1115

\bibitem[{{Lewis}(1974)}]{Lewis74}
{Lewis}, J.~S. 1974, Science, 186, 440

\bibitem[{{Lissauer} {et~al.}(2009){Lissauer}, {Hubickyj}, {D'Angelo}, \&
  {Bodenheimer}}]{Lissauer09}
{Lissauer}, J.~J., {Hubickyj}, O., {D'Angelo}, G., \& {Bodenheimer}, P. 2009,
  \icarus, 199, 338

\bibitem[{{Lodders}(2009)}]{Lodders09}
{Lodders}, K. 2009, ArXiv e-prints

\bibitem[{{Madhusudhan} {et~al.}(2011){Madhusudhan}, {Harrington}, {Stevenson},
  {Nymeyer}, {Campo}, {Wheatley}, {Deming}, {Blecic}, {Hardy}, {Lust},
  {Anderson}, {Collier-Cameron}, {Britt}, {Bowman}, {Hebb}, {Hellier},
  {Maxted}, {Pollacco}, \& {West}}]{Madhusudhan11}
{Madhusudhan}, N., {Harrington}, J., {Stevenson}, K.~B., {et~al.} 2011, Nature,
  469, 64

\bibitem[{{Makalkin} \& {Dorofeeva}(2009)}]{Makalkin09}
{Makalkin}, A.~B. \& {Dorofeeva}, V.~A. 2009, Solar System Research, 43, 508

\bibitem[{{Marois} {et~al.}(2008){Marois}, {Macintosh}, {Barman}, {Zuckerman},
  {Song}, {Patience}, {Lafreni{\`e}re}, \& {Doyon}}]{Marois08}
{Marois}, C., {Macintosh}, B., {Barman}, T., {et~al.} 2008, Science, 322, 1348

\bibitem[{{Mousis} {et~al.}(2009){Mousis}, {Marboeuf}, {Lunine}, {Alibert},
  {Fletcher}, {Orton}, {Pauzat}, \& {Ellinger}}]{Mousis09}
{Mousis}, O., {Marboeuf}, U., {Lunine}, J.~I., {et~al.} 2009, \apj, 696, 1348

\bibitem[{{Pollack} {et~al.}(1996){Pollack}, {Hubickyj}, {Bodenheimer},
  {Lissauer}, {Podolak}, \& {Greenzweig}}]{Pollack96}
{Pollack}, J.~B., {Hubickyj}, O., {Bodenheimer}, P., {et~al.} 1996, Icarus,
  124, 62

\bibitem[{{Pontoppidan}(2006)}]{Pontoppidan06}
{Pontoppidan}, K.~M. 2006, A\&A, 453, L47

\bibitem[{{Richardson} {et~al.}(2006){Richardson}, {Harrington}, {Seager}, \&
  {Deming}}]{Richardson06}
{Richardson}, L.~J., {Harrington}, J., {Seager}, S., \& {Deming}, D. 2006, ApJ,
  649, 1043

\bibitem[{{Siess} {et~al.}(2000){Siess}, {Dufour}, \& {Forestini}}]{Siess00}
{Siess}, L., {Dufour}, E., \& {Forestini}, M. 2000, \aap, 358, 593

\bibitem[{{Stevenson} \& {Lunine}(1988)}]{Stevenson88}
{Stevenson}, D.~J. \& {Lunine}, J.~I. 1988, Icarus, 75, 146

\bibitem[{{Swain} {et~al.}(2010){Swain}, {Deroo}, {Griffith}, {Tinetti},
  {Thatte}, {Vasisht}, {Chen}, {Bouwman}, {Crossfield}, {Angerhausen},
  {Afonso}, \& {Henning}}]{Swain10}
{Swain}, M.~R., {Deroo}, P., {Griffith}, C.~A., {et~al.} 2010, Nature, 463, 637

\bibitem[{{Swain} {et~al.}(2009){Swain}, {Tinetti}, {Vasisht}, {Deroo},
  {Griffith}, {Bouwman}, {Chen}, {Yung}, {Burrows}, {Brown}, {Matthews},
  {Rowe}, {Kuschnig}, \& {Angerhausen}}]{Swain09}
{Swain}, M.~R., {Tinetti}, G., {Vasisht}, G., {et~al.} 2009, ApJ, 704, 1616

\bibitem[{{Tinetti} {et~al.}(2007){Tinetti}, {Vidal-Madjar}, {Liang},
  {Beaulieu}, {Yung}, {Carey}, {Barber}, {Tennyson}, {Ribas}, {Allard},
  {Ballester}, {Sing}, \& {Selsis}}]{Tinetti07}
{Tinetti}, G., {Vidal-Madjar}, A., {Liang}, M., {et~al.} 2007, Nature, 448, 169

\bibitem[{{Visscher} {et~al.}(2010){Visscher}, {Moses}, \&
  {Saslow}}]{Visscher10}
{Visscher}, C., {Moses}, J.~I., \& {Saslow}, S.~A. 2010, Icarus, 209, 602

\bibitem[{{Visser} {et~al.}(2011){Visser}, {Doty}, \& {van
  Dishoeck}}]{Visser11}
{Visser}, R., {Doty}, S.~D., \& {van Dishoeck}, E.~F. 2011, ArXiv e-prints

\bibitem[{{Whittet}(2010)}]{Whittet10}
{Whittet}, D.~C.~B. 2010, ApJ, 710, 1009

\end{thebibliography}

\begin{table*}
\caption{Evaporation temperatures and abundances of O and C in different forms with respect to hydrogen. Adopted model values in parentheses}
\begin{center}
\vspace{-0.3cm}
\begin{tabular}{l c c c }
\hline\hline
Species 	& T$_{\rm evap}^{\rm a}$ & n$_{\rm O}$ & n$_{\rm C}$\\
		&(K)					&(10$^{-4}\times$n$_{\rm H}$) &(10$^{-4}\times$n$_{\rm H}$) \\
\hline
CO		&18--22 (20)		&0.9-2$^{\rm b}$ (1.5)	&0.9-2$^{\rm b}$ (1.5)\\
CO$_2$	&42--52 (47)		&0.6$^{\rm b}$		&0.3$^{\rm b}$		\\
H$_2$O	&120--150 (135)	&0.9$^{\rm b}$		&				\\
Carbon grains	&$>$150 (500)	&		&0.6-1.2$^{\rm c}$ (0.6)\\
Silicate	&$\sim$1500 (1500)	&1.4$^{\rm c}$		&				\\
\hline
\end{tabular}
\label{tab:abund}
\\$^{\rm a}$The range of temperatures for ices corresponds to gas densities 10$^8$ --10$^{12}$ cm$^{-3}$ suitable for disk midplanes.\\
$^{\rm b}$From ice and gas observations toward the CBRR 2422.8-3423 disk \citep{Pontoppidan06}.\\
$^{\rm c}$The range corresponds to estimates of organic content \citep{Draine03}. The lower value is adopted to obtain a solar C/O ratio. Silicate abundance is 1.2 from \citet{Whittet10} and 1.4 takes into account the additional refractory component.
\end{center}
\end{table*}%

\begin{figure}[htp]	
  \centering
 \plotone{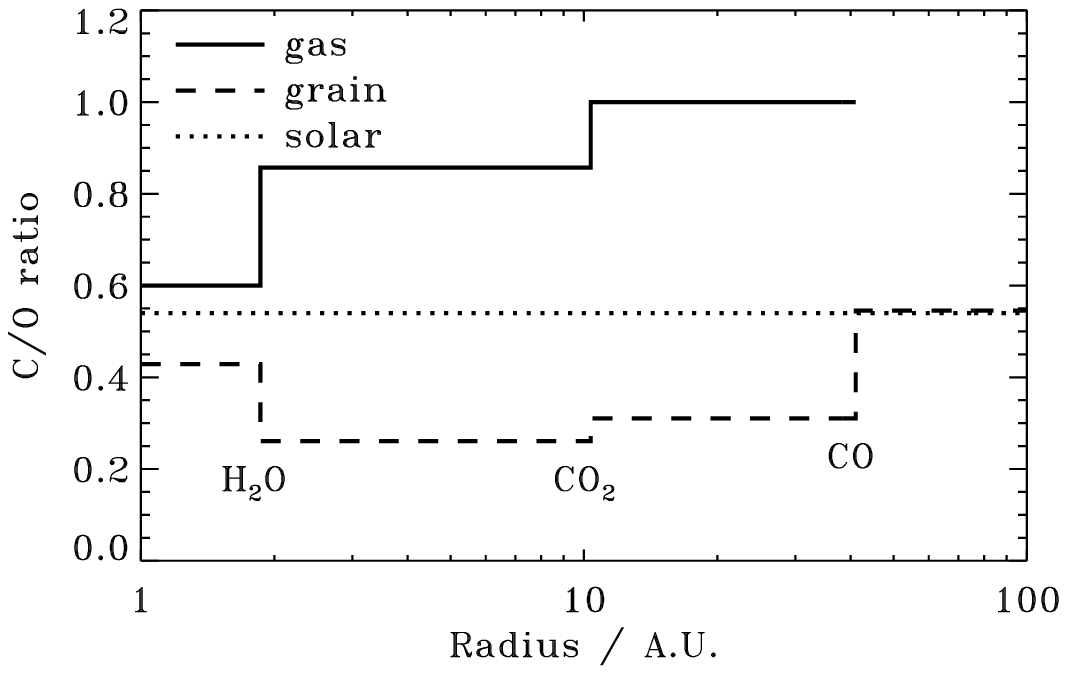}
\caption{The C/O ratio in the gas and in grains, assuming the temperature structure of a `typical' protoplanetary disk around a solar-type star ($T_{\rm 0}$ is 200~K, and $q=0.62$). The H$_2$O, CO$_2$ and CO snow-lines are marked for reference. \label{fig1}}
\end{figure}

\begin{figure}[htp]	
  \plotone{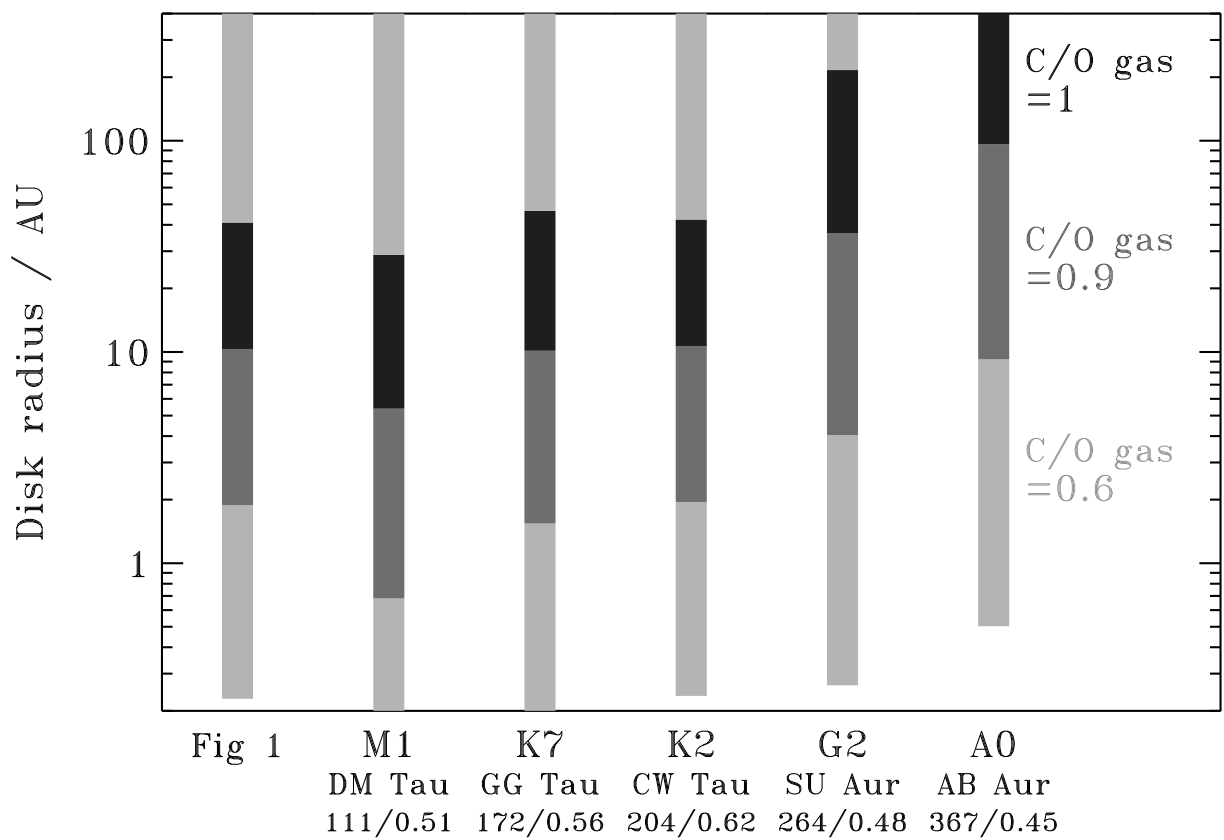}
\caption{The predicted gas phase C/O ratio as a function of radius for five representative disks, ordered by spectral type, compared with the `typical' disk model in Fig. 1. The derived temperature profile parameters, $T_{\rm 0}$ and $q$, are listed. The C/O ratios are calculated assuming that the stellar C/O ratio is solar, i.e. 0.54, and a static disk. \label{fig2}}
\end{figure}

\begin{figure}[htp]	
\plotone{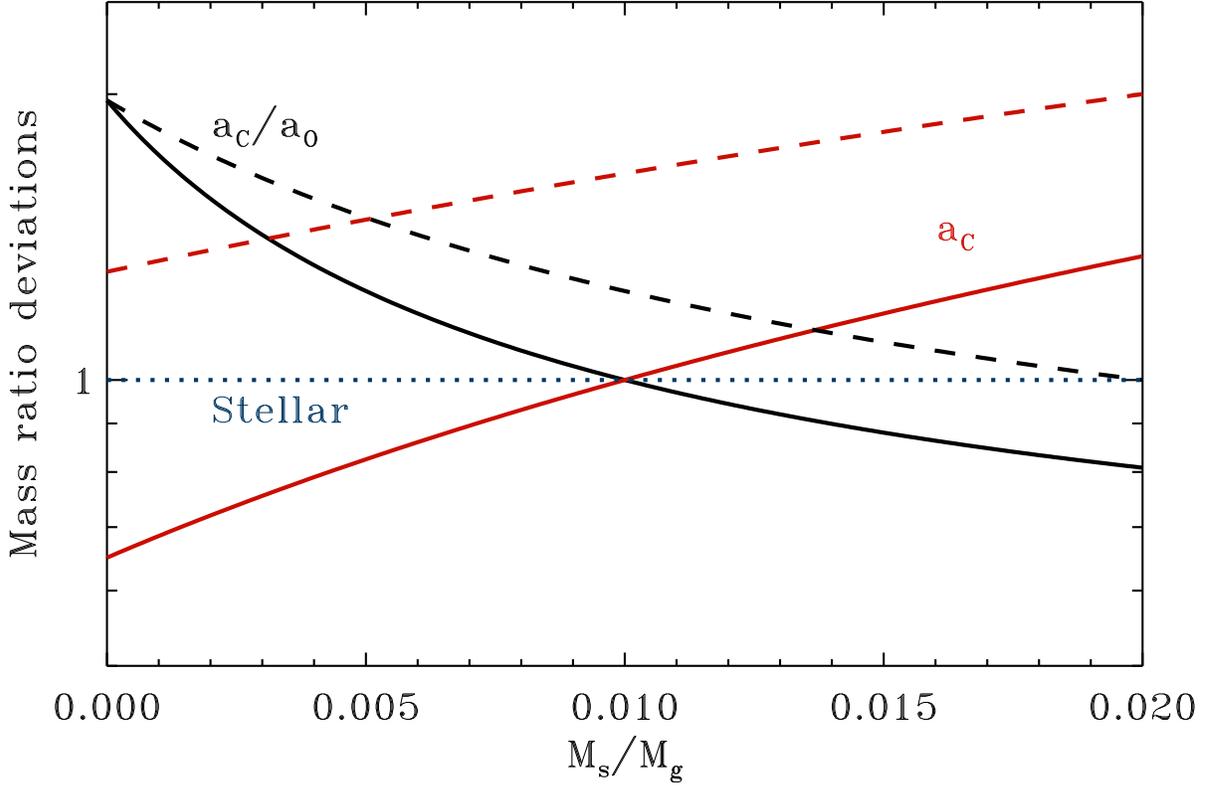}
\caption{The predicted atmospheric deviations of C/H ($a_{\rm C}$, red) and C/O ($a_{\rm C}/a_{\rm O}$, black) from the stellar values, as a function of the fraction of the planet atmosphere mass that comes from planetesimal accretion (as opposed to gas accretion) for a planet forming between the CO$_2$ and CO snowlines. The dashed lines assumed CO/H$_2$ gas enrichment by a factor of 2 because of an evaporation front close to the CO snowline, while the full lines assume the equilibrium CO/H$_2$ gas ratio in a stationary disk. The stationary disk abundance ratios coincide with stellar abundances at 0.01 because this is the assumed grain/gas mass ratio $f_{\rm s/g}$ in the disk. \label{fig3}}
\vspace{4cm}
\end{figure}

\end{document}